\documentclass[preprint,superscriptaddress,showpacs,preprintnumbers,amsmath,amssymb]{revtex4}

\usepackage{graphicx}% Include figure files
\usepackage{dcolumn}% Align table columns on decimal point
\usepackage{bm}% bold math

\begin{document}

\title{Laser-interferometric Detectors for Gravitational Wave Backgrounds at 100 \rm{\bf{MHz}} : Detector Design and Sensitivity}

\author{Atsushi~Nishizawa}
\email{atsushi.nishizawa@nao.ac.jp}
\affiliation{Graduate School of Human and Environmental Studies, Kyoto University, Kyoto 606-8501, Japan}
\author{Seiji~Kawamura}
\affiliation{TAMA Project, National Astronomical Observatory of Japan, Mitaka, Tokyo 181-8588, Japan}
\author{Tomotada~Akutsu}
\affiliation{Department of Astronomy, School of Science, University of Tokyo, Bunkyo-ku, Tokyo 113-0033, Japan}
\author{Koji~Arai}
\affiliation{TAMA Project, National Astronomical Observatory of Japan, Mitaka, Tokyo 181-8588, Japan}
\author{Kazuhiro~Yamamoto}
\affiliation{TAMA Project, National Astronomical Observatory of Japan, Mitaka, Tokyo 181-8588, Japan}
\author{Daisuke~Tatsumi}
\affiliation{TAMA Project, National Astronomical Observatory of Japan, Mitaka, Tokyo 181-8588, Japan}
\author{Erina~Nishida}
\affiliation{Graduate School of Humanities and Sciences, Ochanomizu University, Bunkyo-ku, Tokyo 112-8610, Japan}
\author{Masa-aki~Sakagami}
\affiliation{Graduate School of Human and Environmental Studies, Kyoto University, Kyoto 606-8501, Japan}
\author{Takeshi~Chiba}
\affiliation{Department of Physics, College of Humanities and Sciences, Nihon University, Tokyo 156-8550, Japan}
\author{Ryuichi~Takahashi}
\affiliation{Graduate School of Science, Nagoya University, Nagoya 467-8602, Japan}
\author{Naoshi~Sugiyama}
\affiliation{Graduate School of Science, Nagoya University, Nagoya 467-8602, Japan}

\date{\today}

\begin{abstract}
Recently, observational searches for gravitational wave background (GWB) have been developed and given direct and indirect constraints on the energy density of GWB in a broad range of frequencies. These constraints have already rejected some theoretical models of large GWB spectra. However, at $100 \, \rm{MHz}$, there is no strict upper limit from {\it{direct}} observation, though the {\it{indirect}} limit by $^4\rm{He}$ abundance due to big-bang nucleosynthesis exists.
In this paper, we propose an experiment with laser interferometers searching GWB at $100 \, \rm{MHz}$. We considered three detector designs and evaluated the GW response functions of a single detector. As a result, we found that, at $100 \, \rm{MHz}$, the most sensitive detector is the design, a so-called synchronous recycling interferometer, which has better sensitivity than an ordinary Fabry-Perot Michelson interferometer by a factor of 3.3 at $100 \, \rm{MHz}$. When we select the arm length of $0.75\,\rm{m}$ and realistic optical parameters, the best sensitivity achievable is $h \approx 7.8 \times 10^{-21}\,\rm{Hz}^{-1/2}$ at $100 \, \rm{MHz}$ with bandwidth $\sim 2 \, \rm{kHz}$.
\end{abstract}
\maketitle

\section{Introduction}
There are many theoretical predictions of gravitational wave background (GWB) in a broad range of frequencies, $10^{-18}-10^{10}\, \rm{Hz}$. Some models in cosmology and particle physics predict that there are large stochastic
GWB at ultra high frequency $\sim 100 \, \rm{MHz}$. Testing these models with GW detectors for
high frequencies is very important. In the quintessential inflation model \cite{bib10}, the blue spectrum of GWB is produced during the kinetic energy-dominated era after the inflationary expansion of the universe \cite{bib11,bib12,bib13} and is very sensitive to subsequent reheating processes \cite{bib14}. In other inflation models, during the first stage of the process of reheating, called preheating, GWB at high frequencies is created due to large density inhomogeneities \cite{bib29,bib30,bib34}.
Pre-big-bang scenarios in string cosmology can also generate high-frequency backgrounds \cite{bib15,bib16,bib17}. Not only cosmological sources but also compact objects can create GWB at $\sim 100 \, \rm{MHz}$. Primordial black holes produced in the early universe, which have much smaller masses than the sun, emit GW via binary evolution and coalescence \cite{bib18,bib19} and evaporation \cite{bib20}. Recent predictions of GW emission from black strings in the Randall-Sundrum model generate spectral features characteristic of the curvature of extra dimensions at high frequencies \cite{bib21,bib22}. 

Upper limits on GWB in wide-frequency ranges have been obtained from various observations; cosmic microwave radiation at $10^{-18}-10^{-15} \,\rm{Hz}$ \cite{bib23}, pulsar timing at $10^{-9}-10^{-7}\, \rm{Hz}$ \cite{bib24}, Doppler tracking of the {\it{Cassini}} spacecraft at $10^{-6}-10^{-3}\, \rm{Hz}$ \cite{bib33}, direct observation by LIGO at $10-10^{4} \rm{Hz}$ \cite{bib25}, $^4\rm{He}$ abundance due to big-bang nucleosynthesis at greater frequencies than $10^{-10}\, \rm{Hz}$ \cite{bib26}. Nevertheless, as far as we know, no direct experiment has been done above $10^5 \, \rm{Hz}$ except for the experiment by A. M. Cruise and R. M. J. Ingley \cite{bib27}. They have used electromagnetic waveguides and obtained an upper limit on the amplitude of GW backgrounds, $h\leq 10^{-14}$ corresponding to $ h_{100}^2\Omega_{\rm{gw}} \leq 10^{34}$ at $100 \, \rm{MHz}$, where $h_{100}$ is the Hubble constant normalized with $100 \,\rm{km}\,\rm{sec}^{-1}\,\rm{Mpc}^{-1}$ and $\Omega_{\rm{gw}}$ is the energy density of GWB per logarithmic frequency bin normalized by the critical energy density of the universe, that is $\Omega_{\rm{gw}}(f) = (d\rho_{\rm{gw}}/d \ln f)/ \rho_c$. This constraint is much weaker than the constraints at other frequencies. Therefore, a much tighter bound above $10^5 \, \rm{Hz}$ is needed to test various theoretical models.

In this paper, we propose a method of direct detection of GW at $100 \, \rm{MHz}$ with laser interferometers. At high frequencies, the GW wavelength is comparable to the size of a detector, which is the order of a few meters. The usual approximation that the GW wavelength is much larger than the detector size is not valid. In this case, the phase of GW changes during the one-way trip of light between mirrors. Therefore, we have to use a detector design that is able to integrate GW signals efficiently. Note that our investigation is general and applicable to other than $100 \, \rm{MHz}$. The contents of the paper are as follows: In Sec. II, we will consider three detector designs and give GW response functions. Detailed calculations are given in the Appendices. In Sec. III, the GW responses numerically calculated are compared. Sec. IV is devoted to discussions and conclusions.   

\section{Detector Designs and the response functions}
In this section, we will consider three detector designs, (i) synchronous recycling interferometer (SRI), (ii) Fabry-Perot Michelson interferometer (FPMI) and (iii) L-shaped cavity Michelson interferometer (LMI), and derive the response functions for GWs.

\subsection{Detector designs}
\label{sec1}
\begin{figure}[h]
\begin{center}
\includegraphics[width=8cm]{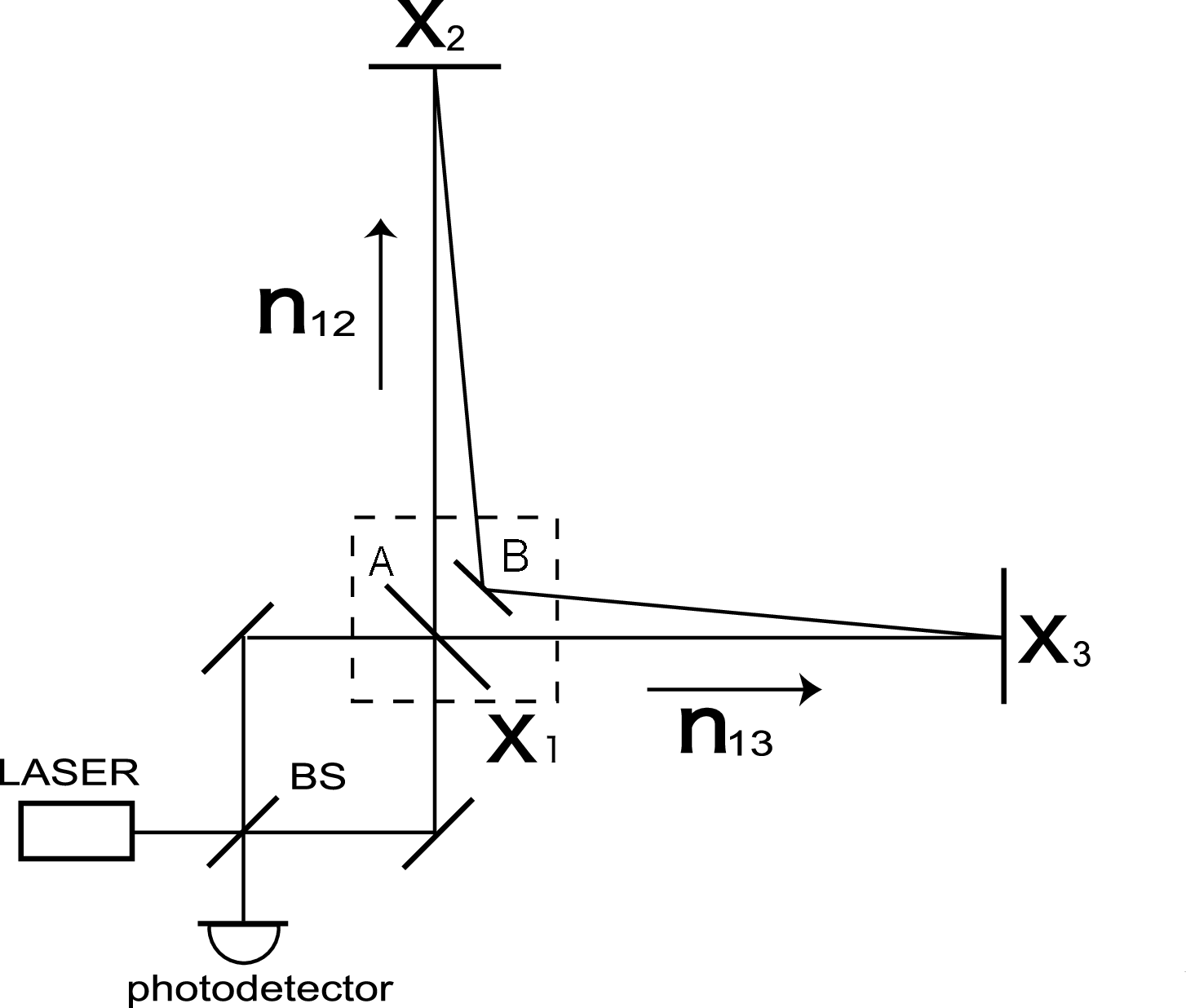}
\caption{Synchronous recycling interferometer(SRI).}
\label{fig1}
\end{center}
\end{figure}

SRI (Fig.\ref{fig1}) was first proposed by R. W. P. Drever in \cite{bib1} and detailed calculations have been done in \cite{bib31,bib32}. Laser light is split at a beam splitter and sent into an SR cavity through a recycling mirror, which is mirror A located at $\mathbf{X}_1$ in Fig.\ref{fig1}. The beams circulating clockwise and counterclockwise in the cavity experience gravitational waves and mirror displacements, leave the cavity, and are recombined at the beam splitter. Then, the differential signal is detected at a photodetector. The advantage of SRI is that GW signals at certain frequencies are accumulated and amplified because the light beams experience GWs with the same phases during round trips in the folded cavity.
Consider GW propagating normally to the detector plane with an optimal polarization. In this case, the GW signal is amplified at the frequencies $f = (2n-1) \times c/4L,\;n=1,2,\cdots $, where $c$ is the speed of light and $L$ is the arm length. More precisely, the arm length is the distance between $\mathbf{X}_1$ and $\mathbf{X}_2$ (or $\mathbf{X}_3$) in Fig.\ref{fig1}. On the other hand, the disadvantage of SRI is less sensitivity for GWs at low frequencies, $f < c/4L$, because the GW signal is integrated in the cavity and canceled out as the frequency is low. 

\begin{figure}[h]
\begin{center}
\includegraphics[width=8cm]{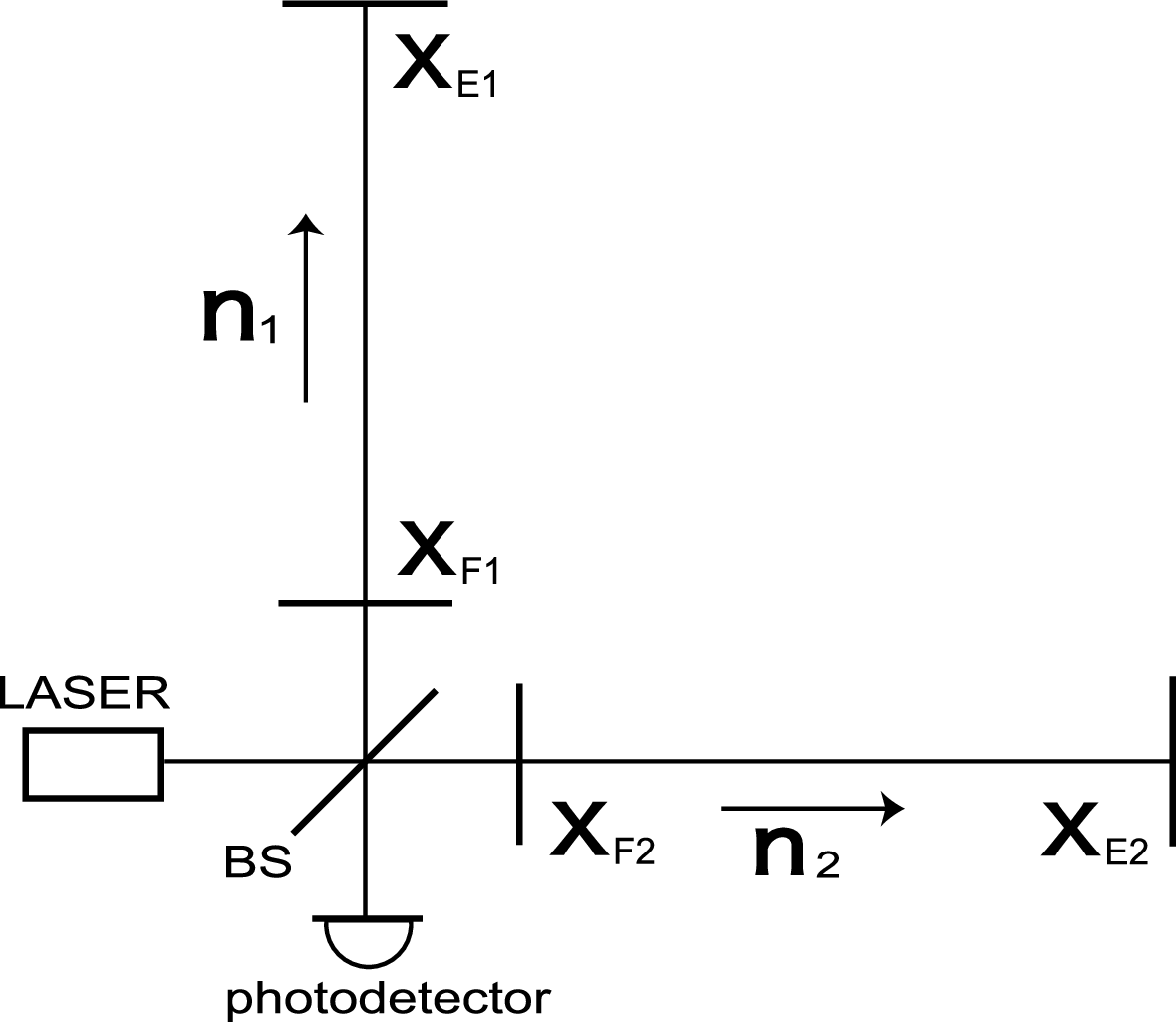}
\caption{Fabry-Perot Michelson interferometer(FPMI).}
\label{fig2}
\end{center}
\end{figure}

The competitive design of detectors with SRI is an ordinary FPMI (Fig.\ref{fig2}). FPMI is the most popular design for current ground-based interferometers \cite{bib2,bib3,bib4,bib5} since it has good sensitivity at low frequencies due to the amplification of GW signals with Fabry-Perot cavities. However, to amplify the GW signals at high frequencies, one needs to use resonance due to the cavity. FPMI has the resonance of GW signals at the frequencies, $f = n \times c/2L,\; n=1,2,\cdots$ when the GW response is averaged over the entire sky. To take advantage of the resonant response to GW at $100\rm{MHz}$, the arm length of FPMI should be $1.5\, \rm{m}$. With this detector, one can achieve good sensitivity with narrow bandwidth as well as SRI.

\begin{figure}[h]
\begin{center}
\includegraphics[width=8cm]{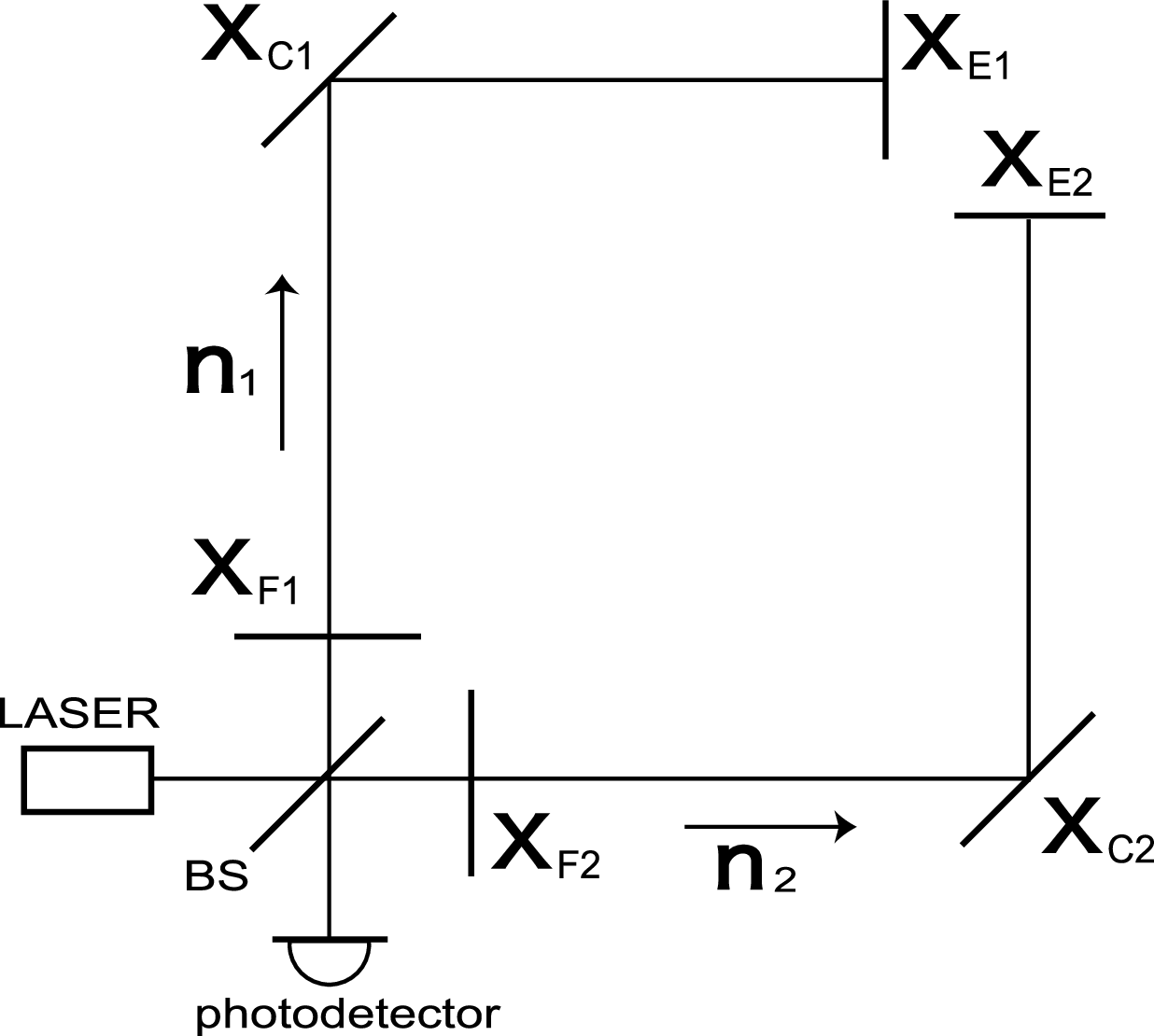}
\caption{L-shaped cavity Michelson interferometer(LMI).}
\label{fig3}
\end{center}
\end{figure}

Another possible design of detectors is LMI (Fig.\ref{fig3}), whose optical configuration is the same as the L-shaped FPMI. However, the GW response resembles SRI rather than FPMI. Thus, this design can be regarded as being intermediate between SRI and FPMI.

To compare the detectors for GWB at $100\, \rm{MHz}$, it is necessary to derive the GW response functions for GW propagating in arbitrary directions and to compare those averaged on the celestial sphere. For GW propagating normal to the detector plane with an optimal polarization, it is trivial that SRI and LMI with arm length $0.75\, \rm{m}$ have maximal sensitivity, while FPMI with $1.5\, \rm{m}$ is not sensitive at all, at $100\, \rm{MHz}$. However, FPMI has nonzero sensitivity for GW not propagating orthogonally to the arms of the detector
since the GW response of light {\it{going}} and {\it{coming}} differs. Furthermore, the geometries of detectors also affect the GW responses. Therefore, it is nontrivial which is the most sensitive detector.
  
\subsection{GW response functions}
The general expression of the phase shift of light induced by GW propagating in an arbitrary direction has been derived in many references, for example, \cite{bib6,bib7}. We will use the expression in \cite{bib6} as our starting point. When light travels between two test masses located at $\mathbf{X}_i$ and $\mathbf{X}_j$, the phase shift of light created by GW in TT (transverse-traceless) gauge is expressed as
\begin{equation}
\delta \phi_{ij}(t)=(\mathbf{n}_{ij} \otimes \mathbf{n}_{ij}) :  \frac{\omega}{2} \int _0^{L_{ij}/c}dt^{\prime} \sum _{p} \mathbf{e}^p h_p 
[ t-(L_{ij}+\mathbf{e}_z \cdot \mathbf{X}_i)/c + (1-\mathbf{e}_z \cdot \mathbf{n}_{ij})t^{\prime} ],
\label{eq1}
\end{equation}
where $t$ is the reception time of light at $\mathbf{X}_j$, $\omega$ is the angular frequency of light, $h_p$ is the amplitude of GW with plus or cross polarization and $\mathbf{e}_z$ is a unit vector in the direction of GW propagation. The arm length and unit vector in the direction of the arm are defined as $L_{ij} \equiv |\mathbf{X}_j-\mathbf{X}_i|$ and $\mathbf{n}_{ij}\equiv (\mathbf{X}_j-\mathbf{X}_i)/L_{ij}$, respectively. The symbol : means contraction between a tensor and vectors. The tensors $\mathbf{e}_p, \;p=+,\times$ are polarization tensors of GW and are defined as
\begin{eqnarray}
\mathbf{e}_{+} &\equiv & \mathbf{e}_x \otimes \mathbf{e}_x -\mathbf{e}_y \otimes \mathbf{e}_y \;,
\label{eq5} \\
\mathbf{e}_{\times} &\equiv & \mathbf{e}_x \otimes \mathbf{e}_y +\mathbf{e}_y \otimes \mathbf{e}_x \;, 
\label{eq6}
\end{eqnarray}
where $\mathbf{e}_x$ and $\mathbf{e}_y$ are the unit vectors, which form the orthogonal coordinate with $\mathbf{e}_z$. We assume that there is no displacement noise, for example, thermal noise, seismic noise, radiation pressure noise, etc., at $100\, \rm{MHz}$. (In fact, rough estimates show that this assumption is at least valid as long as the detector sensitivity is above $h \sim 10^{-20}\, \rm{Hz}^{-1/2}$. In the experiment that reaches better sensitivity, thermal noise of mirrors might limit the sensitivity, though other noises are far below. In that case, one should calculate the noise spectrum accurately with numerical simulation including the effect of the mirror's complicated response at $100\, \rm{MHz}$, which is beyond the scope of this paper.) Under the assumption of no displacement, the positions of mirrors in the absence of GW are not perturbed and are just given by $\mathbf{X}_i$ and $\mathbf{X}_j$. The Fourier transform of Eq. (\ref{eq1}) is given by 
\begin{eqnarray}
\tilde{\delta \phi}_{ij}(\Omega )&=& \mathbf{n}_{ij} \otimes \mathbf{n}_{ij} \; : \; 
\frac{\omega}{2} \, \sum_p \mathbf{e}^p \tilde{h}_p
\frac{e^{-i\Omega (\mathbf{e}_z \cdot \mathbf{X}_j)/c}
-e^{-i\Omega (L_{ij}+\mathbf{e}_z \cdot \mathbf{X}_i)/c}}
{i\Omega (1-\mathbf{e}_z \cdot \mathbf{n}_{ij})}\, ,
\label{eq2}
\end{eqnarray}
where $\tilde{h}_p$ is the Fourier component of the GW amplitude, and $\Omega$ is the angular frequency of GW and is related to the GW frequency with $\Omega = 2\pi f$.

In general, the response function of a detector is represented by the round-trip signal in cavities multiplied by an amplification factor in cavities. We denote the phase shift of
the round-trip signal by $\tilde{\delta \phi} (\Omega)$ and the amplification factor by $\alpha (\Omega )$. Then, the total output from the detector $\tilde{\delta \Phi} (\Omega)$ is written as $\tilde{\delta \Phi }(\Omega)=\alpha (\Omega ) \tilde{\delta \phi} (\Omega)$.
Detailed calculations are described in the Appendices. We show here only the results. Note that we change the notation of the unit vectors directed in arms and the reflectivity of mirrors in order to simplify the expression and make it easy to compare. The response functions of each detector are
\begin{equation}
\tilde{\delta \Phi}_{all} (\Omega )= \alpha (\Omega, R_F,R_E) \tilde{\delta \phi}_{all} (\Omega)\;,
\label{eq66}
\end{equation}
\begin{equation}
\alpha (\Omega, R_F,R_E) =-\frac{R_E T_F^2}{(R_F-R_E)(1-R_F R_E \;e^{-4i\Omega \tau})}\;, \label{eq3}
\end{equation}
and
\begin{itemize}
\item{SRI}\\
Replacing $\mathbf{n}_{12}\rightarrow \mathbf{u}$, $\mathbf{n}_{13}\rightarrow \mathbf{v}$ and $(R_A, R_C)\rightarrow (R_F, R_E)$ in Eq. (\ref{eq63}), the response function is
\begin{eqnarray}
\tilde{\delta \phi}_{all}(\Omega ) &=& (1-e^{-2i\Omega \tau})  \frac{\omega}{i\Omega}\;e^{-i\Omega (\tau+\mathbf{e}_z \cdot \mathbf{X}_1/c)} \sum_p \mathbf{e}^p \tilde{h}_p \nonumber \\
&&: \left[ \right. \frac{\mathbf{v} \otimes \mathbf{v}}{1-(\mathbf{e}_z \cdot \mathbf{v} )^2 } \{ i \;\sin\Omega \tau +(\mathbf{e}_z \cdot \mathbf{v})(e^{-i\Omega \tau \mathbf{e}_z \cdot \mathbf{v}}-\cos\Omega \tau) \} \nonumber \\
&&\;-\frac{\mathbf{u} \otimes \mathbf{u}}{1-(\mathbf{e}_z \cdot \mathbf{u})^2} \{ i \;\sin\Omega \tau +(\mathbf{e}_z \cdot \mathbf{u})(e^{-i\Omega \tau \mathbf{e}_z \cdot \mathbf{u}}-\cos\Omega \tau) \} \left. \right]\;,
\end{eqnarray}
\item{FPMI (doubled the arm length, $\tau \rightarrow 2\tau$)}\\
Replacing $\mathbf{n}_{1}\rightarrow \mathbf{u}$, $\mathbf{n}_{2}\rightarrow \mathbf{v}$ in Eq. (\ref{eq64}), the response function is
\begin{eqnarray}
\tilde{\delta \phi}_{all}(\Omega ) &= & \frac{\omega}{i\Omega} \sum_{p} \mathbf{e}^p \tilde{h}_p \;e^{-i\Omega (2\tau+\mathbf{e}_z \cdot \mathbf{X}_F/c)} \nonumber \\
&&: \left[ \right. \frac{\mathbf{u} \otimes \mathbf{u}}{ 1-(\mathbf{e}_z \cdot \mathbf{u})^2 } 
\left\{ \right. i\;\sin2\Omega \tau +(\mathbf{e}_z \cdot \mathbf{u})(e^{-2i\Omega \tau \mathbf{e}_z \cdot \mathbf{u}}-\cos2\Omega \tau) \left. \right\} \nonumber \\
&&-\frac{\mathbf{v} \otimes \mathbf{v}}{ 1-(\mathbf{e}_z \cdot \mathbf{v})^2 } 
\left\{ \right.  i\;\sin2\Omega \tau +(\mathbf{e}_z \cdot \mathbf{v})(e^{-2i\Omega \tau \mathbf{e}_z \cdot \mathbf{v}}-\cos2\Omega \tau) \left. \right\} \left. \right]\;,
\end{eqnarray}
\item{LMI}\\
Replacing $\mathbf{n}_{1}\rightarrow \mathbf{u}$, $\mathbf{n}_{2}\rightarrow \mathbf{v}$ in Eq. (\ref{eq65}), the response function is
\begin{eqnarray}
\tilde{\delta \phi}_{all}(\Omega ) &=&  \frac{\omega}{i\Omega} \sum_{p} \mathbf{e}^p \tilde{h}_p \;e^{-i\Omega (2\tau+\mathbf{e}_z \cdot \mathbf{X}_{F}/c)} \nonumber \\
&&: \left[ \right. \frac{\mathbf{v}\otimes \mathbf{v}}{ 1-(\mathbf{e}_z \cdot \mathbf{v})^2 } \left\{ \right. i\;\sin\Omega \tau\; (e^{-ip_1}+e^{-ip_2}-2\cos\Omega \tau) \nonumber \\
&&\;\;\;\;\;\;\;\;\;\;\;\;\;\;\;\;\;\;\;\;\;\;+(\mathbf{e}_z \cdot \mathbf{v})(e^{-i(p_1+p_2)}+\cos2\Omega \tau -(e^{-ip_1}+e^{-ip_2})\cos\Omega \tau) \left. \right\} \nonumber \\
&&-\frac{\mathbf{u}\otimes \mathbf{u}}{ 1-(\mathbf{e}_z \cdot \mathbf{u})^2 } \left\{ \right. i\;\sin\Omega \tau\; (e^{-ip_1}+e^{-ip_2}-2\cos\Omega \tau) \nonumber \\
&&\;\;\;\;\;\;\;\;\;\;\;\;\;\;\;\;\;\;\;\;\;\;+(\mathbf{e}_z \cdot \mathbf{u})(e^{-i(p_1+p_2)}+\cos2\Omega \tau -(e^{-ip_1}+e^{-ip_2})\cos\Omega \tau) \left. \right\} \left. \right]\;. \nonumber \\
&&
\end{eqnarray} 
\end{itemize}
The phases $p_1$ and $p_2$ are defined by $p_1 \equiv \Omega \tau \;(\mathbf{e}_z \cdot \mathbf{u})$ and $p_2 \equiv \Omega \tau\; (\mathbf{e}_z \cdot \mathbf{v})$, $\tau$ is defined by $\tau \equiv L/c$ and $R_F$ and $R_E$ are the amplitude reflectivities of front and end mirrors of cavities, respectively. $T_F$ is the amplitude transmissivity of a front mirror of cavities.
Note that, in the case of SRI, the front and end mirrors correspond to a recycling mirror and three other mirrors. Here we doubled the arm length of FPMI so that the first resonant frequency of GW signal coincides with that of SRI and LMI.

\section{Detector comparison}
As mentioned in the previous section, in general, the GW response function has the form $\tilde{\delta \Phi }_{all}=\alpha \; \tilde{\delta \phi}_{all} $. We will consider $\alpha$, which is the common factor for all detectors, and $\tilde{\delta \phi}$, which depends on the geometry of each detector, separately.

\begin{figure}[h]
\begin{center}
\includegraphics[width=10cm]{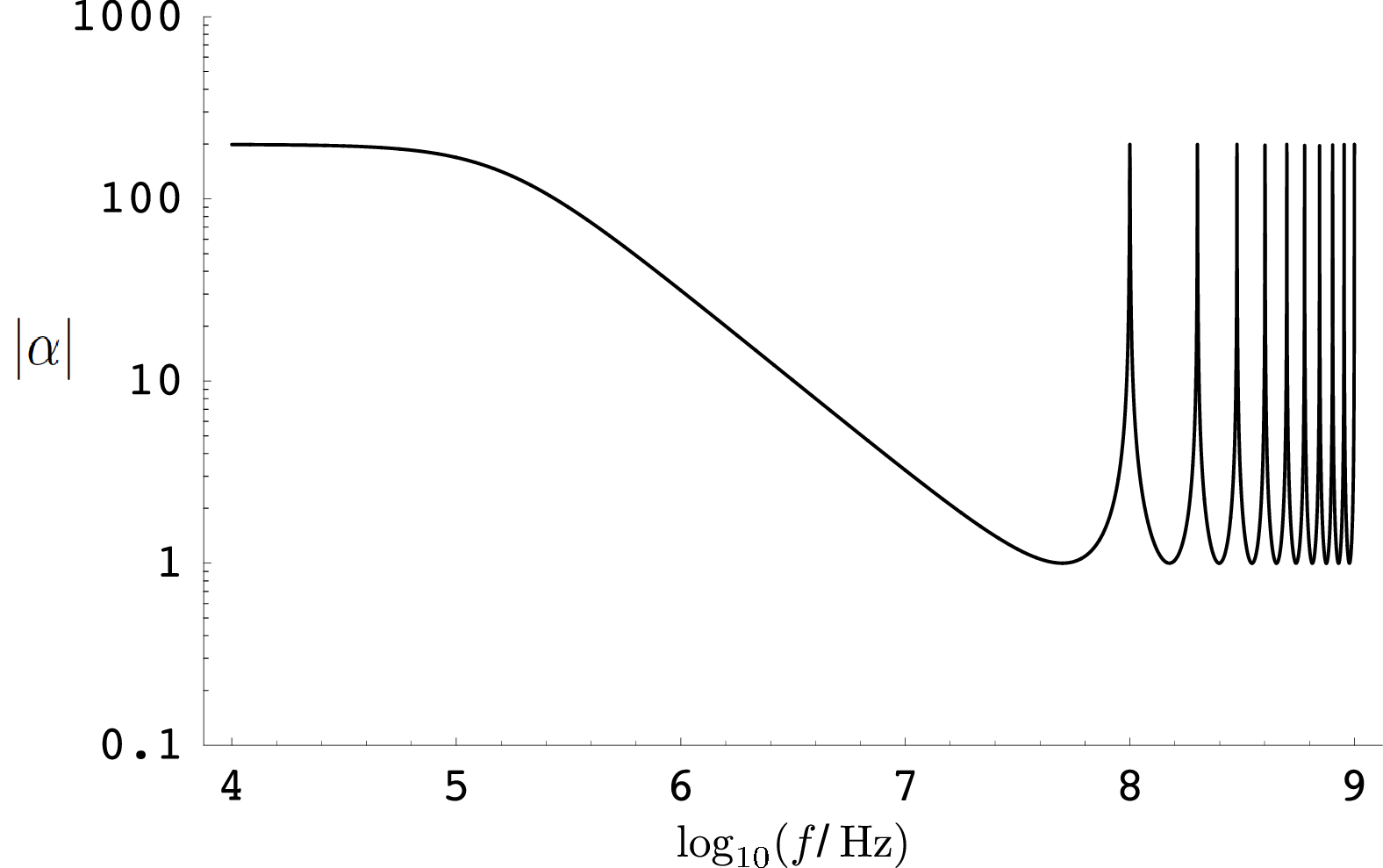}
\caption{The optical amplification factor $\alpha(\Omega)$. Parameters are selected $L=0.75\rm{m}$, $R_F=0.99$ and $R_E=1$.}
\label{fig4}
\end{center}
\end{figure}

From Eq. (\ref{eq3}), the magnitude of the optical amplification factor in the cavities is determined only by the (composite) reflectivities of the front and end mirrors. The frequencies of the peaks depend on the arm length of detectors.
$\alpha$ is plotted in Fig.{\ref{fig4}}. In the figure, we selected $L=0.75\, \rm{m}$ so that the first resonant peak is located at $100\, \rm{MHz}$. At higher frequencies, there are many resonant peaks. At lower frequencies, optical amplification is stronger as the frequency is lower, since the wavelength of light is larger than the arm length of a detector.
Here we selected the amplitude reflectivities $R_F=0.99$ and $R_E=1$ for an illustrative purpose. However, in a real experiment, one should select the reflectivities of the front mirrors much higher in order to achieve better sensitivity, though the bandwidth becomes narrower.

To evaluate the round-trip phase shift due to GWs $\tilde{\delta \phi}_{all}$, we introduce coordinates here. The detectors are located on the X-Y plane. Two unit vectors $\mathbf{u}$ and $\mathbf{v}$ are written as $\mathbf{u}=(0,1,0)$ and $\mathbf{v}=(1,0,0)$, respectively. We denote the direction of GW propagation by the unit vector $\mathbf{e}_z$, and the two unit vectors normal to $\mathbf{e}_z$ and orthogonal to each other by $\mathbf{e}_x$ and $\mathbf{e}_y$. These are written as
\begin{eqnarray}
\mathbf{e}_{x}&=&(\;\cos\theta \;\cos \phi ,\; \cos\theta \;\sin\phi,  \;-\sin\theta \;)\;, \nonumber \\
\mathbf{e}_{y}&=&(\;-\;\sin \phi ,\; \cos\phi ,\;0\;)\;,  \nonumber \\
\mathbf{e}_{z}&=&(\;\sin\theta \;\cos \phi ,\;\sin\theta \;\sin\phi , \;\cos\theta \;)\;. \nonumber 
\end{eqnarray}
$\mathbf{e}_x$ and $\mathbf{e}_y$ define the GW polarization tensor in Eqs. (\ref{eq5}) and (\ref{eq6}). Here we normalize and redefine the GW response function as a dimensionless response function, namely,
\begin{equation}
{\cal{T}}(\Omega ,\phi, \theta , \psi) \equiv \frac{\tilde{\delta \phi}_{all}}{(\omega \tilde{h} \tau )}\;,
\label{eq67}
\end{equation}
where we assumed that GW has the form $\sum \mathbf{e}_p \tilde{h}^p =\tilde{h}\,(\mathbf{e}_+\;\cos2\psi +\mathbf{e}_{\times}\;\sin2\psi )$. $\psi$ is the polarization angle of GW. Integrating this function by $\phi$, $\theta$ and $\psi$ on the celestial sphere and averaging lead to
 \begin{equation}
{\cal{T}}_{rms}^2(\Omega ) \equiv \frac{1}{4\pi } \int_{0}^{2\pi}d\phi \int_{0}^{\pi}d\theta \;\sin \theta \int_{0}^{2\pi} \frac{d\psi}{2\pi} \;|{\cal{T}}(\Omega ,\phi, \theta, \psi )|^2.
\end{equation}

\begin{figure}[t]
\begin{center}
\includegraphics[width=10cm]{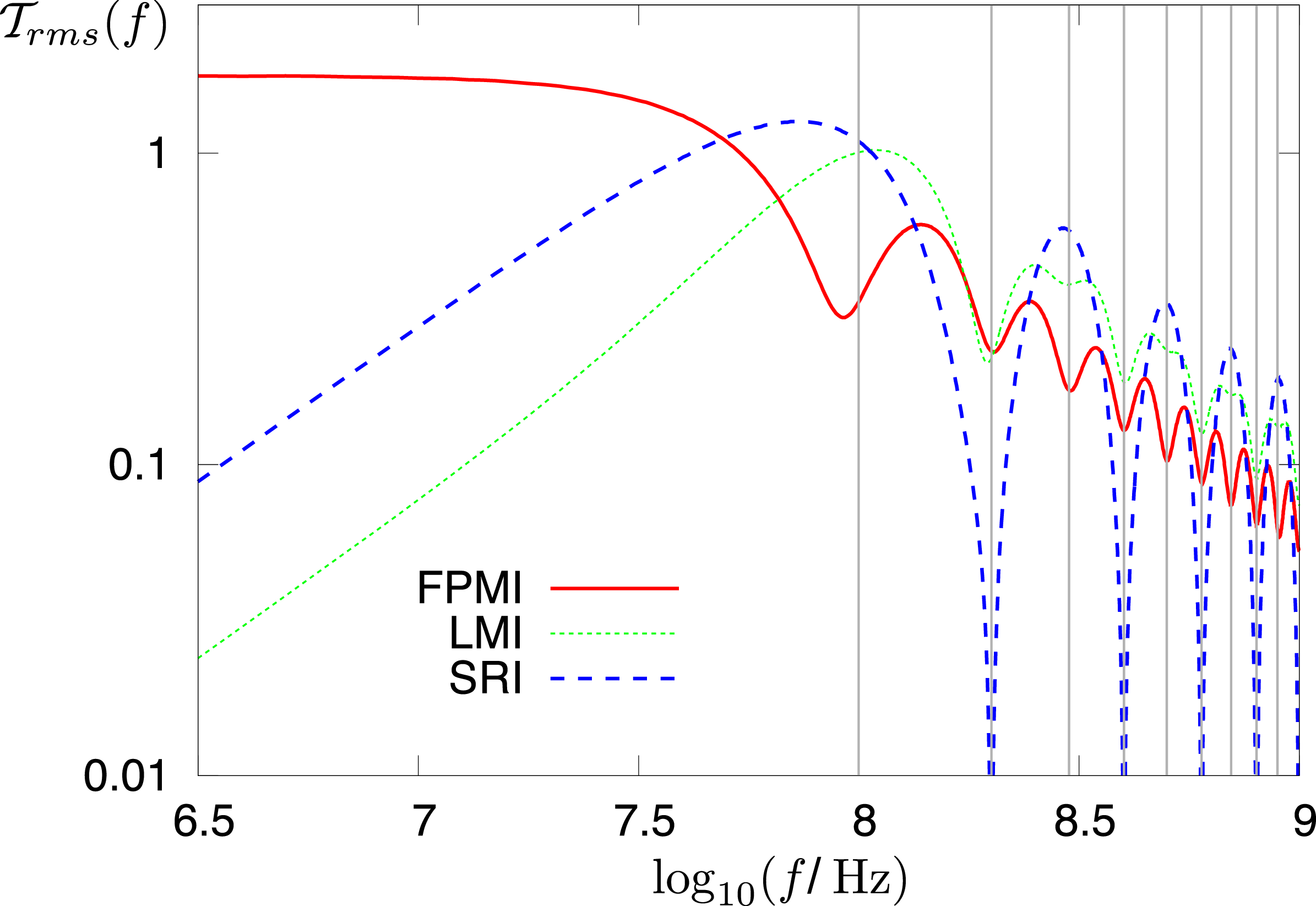}
\caption{(Color online). GW response function ${\cal{T}}_{rms}(f)$. Vertical lines are plotted at multiples of $100\rm{MHz}$.}
\label{fig5}
\end{center}
\end{figure}

The result of numerical calculation is shown in Fig.\ref{fig5}. All response functions decrease above $100\, \rm{MHz}$. This is the common feature of interferometers because the GW signal is destructively integrated in the cavity and is canceled out. Below $100\, \rm{MHz}$, the response functions of SRI and LMI also decrease because the GW signal is partially canceled out during round trips of light beams in the folded cavities. On the other hand, in the case of FPMI, the GW signal is more constructively integrated in the cavity and is more amplified, as the frequency gets lower. At $100\, \rm{MHz}$, SRI and LMI have almost the same sensitivity, while FPMI has sensitivity worse by a factor $\approx 3.3$. This is because FPMI integrates GW signals less efficiently than other detectors, as mentioned in Sec. \ref{sec1}. This difference becomes significant when one takes the correlation of two detectors into account, which results in a factor of $(3.3)^2 \sim 10$ in sensitivity to GWB energy density $\Omega_{\rm{gw}}$.

At the end of this section, let us consider the best sensitivity of SRI experimentally achievable with realistic parameters, which is almost the same as that of LMI. We assume that the sensitivity is limited only by shot noise. The magnitude of shot noise is determined by laser power, the arm length and the reflectivities of mirrors, and is calculated by the condition that the phase shift due to GWs is equal to that of quantum noise \cite{bib31},
\begin{equation}
|\beta |\, \langle \tilde{\delta \Phi}_{all} \rangle _{rms} = \sqrt{\frac{2\hbar \omega}{\eta I_0}}\,,
\label{eq68}
\end{equation}
where $\hbar$ is the reduced Planck constant, $\omega$ is the angular frequency of laser, $\eta$ is the quantum efficiency of a photodetector, $I_0$ is original laser power, and $\beta \equiv (R_F-R_E)/(1-R_F R_E)$. The reason why $\beta$ is needed in the left-hand side of the Eq.(\ref{eq68}) is that, in Eq.(\ref{eq69}), the phase shift due to GWs must be converted into the amplitude of a sideband field.
Substituting Eqs.(\ref{eq66}) and (\ref{eq67}) leads to
\begin{equation}
h (\Omega )= \frac{1}{\tau \alpha ^{\prime}(\Omega ) {\cal{T }}_{rms}(\Omega ) } \sqrt{\frac{2\hbar}{\eta \omega I_0}}\,,
\end{equation}
where we defined
\begin{equation}
\alpha ^{\prime} (\Omega ) \equiv |\beta \,\alpha (\Omega )|= \left| \frac{R_E T_F^2}{(1-R_F R_E)(1-R_F R_E \;e^{-4i\Omega \tau})} \right|\;.
\end{equation}
We select $L=0.75\, \rm{m}$ so that GW signal resonates at $100\, \rm{MHz}$, $\omega =1.77\times 10^{15}\, \rm{rad\: sec}^{-1} $ and $\eta =1$. The sensitivity achievable at $100 \, \rm{MHz}$ in an ideal situation is 
\begin{equation}
h \approx 7.8 \times 10^{-21} \biggl( \frac{1\,\rm{W}}{I_0} \biggr)^{1/2} 
\biggl( \frac{1.6 \times 10^4}{\alpha ^{\prime}} \biggr) \,\rm{Hz}^{-1/2}
\end{equation}
with bandwidth $\sim 10^8/{\cal{F}} \: \rm{Hz}$, where finesse is ${\cal{F}} \equiv \pi \sqrt{R_F R_E}/(1-R_F R_E)$, which is related to $\alpha ^{\prime}$ by the relation 
${\cal{F}} =\pi \sqrt{\alpha ^{\prime} R_F/T_F^2}$ at $100 \, \rm{MHz}$. Note that $\alpha ^{\prime} \approx 1.6 \times 10^4$ is realized with reflectivities, say, $R_F^2=0.99996$ and $R_E^2=(0.99998)^3$.

\section{Discussions and conclusions}
In real experiments, there are many advantages and disadvantages of detectors that we have not considered in this paper. One of the advantages of SRI is the simplicity of a control system. SRI has only one degree of freedom for locking the interferometer because clockwise and counterclockwise lights share light paths in the cavity and the Sagnac part, while FPMI and LMI have three degrees of freedom in operation, which are for cavities in both arms and the Michelson part. Another advantage of SRI is the symmetric optical configuration of the cavity. This means that clockwise and counterclockwise light in the cavity experience the same reflectivities of mirrors. Thus, SRI is expected to have high tolerance to the imbalance of the reflectivities and relatively smaller laser frequency noise than other detectors. Considering these facts, we can conclude that SRI is the best detector. These issues should be investigated in more detail when one constructs real detectors.

In this paper, we investigated the GW responses of interferometers at $100 \, \rm{MHz}$. We considered three designs that took advantage of the first optical resonance due to cavities and derived the GW response functions. As a result, SRI and LMI have almost the same sensitivity at $100 \, \rm{MHz}$ and FPMI has sensitivity worse by a factor of 3.3. This is a significant difference because the sensitivity is better by a factor $(3.3)^2 \sim 10$ in sensitivity to GWB energy density $\Omega_{\rm{gw}}$ when we take the correlation between two detectors into account. When we select the arm length $L=0.75 \, \rm{m}$, laser power $I_0= 1 \, \rm{W}$ and $|\alpha ^{\prime}| = 1.6\times 10^4$, the best achievable sensitivity with SRI is $h \approx 7.8 \times 10^{-21}\, \rm{Hz}^{-1/2}$ at $100 \, \rm{MHz}$ with bandwidth $\sim 2 \, \rm{kHz}$. Note that our results can also be applied to detectors at other frequency bands by tuning the arm length and shifting the peak of sensitivity.

\begin{acknowledgments}
This research was supported by the Ministry of Education, Science, Sports and Culture, Grant-in-Aid for Scientific Research (A), 17204018.
\end{acknowledgments}

\appendix

\section{Response function of SRI}
\begin{figure}[h]
\begin{center}
\includegraphics[width=6cm]{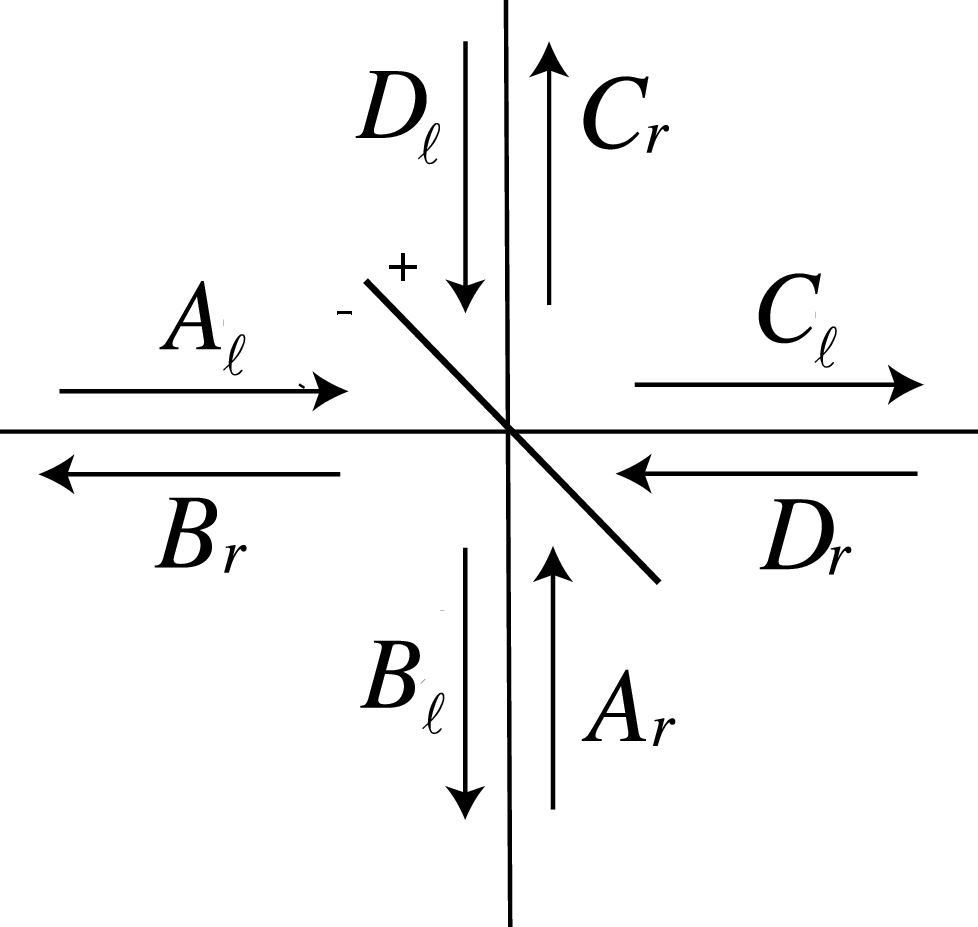}
\caption{Electric fields of SRI at the recycling mirror.}
\label{fig11}
\end{center}
\end{figure}

The configuration of SRI is shown in Fig.\ref{fig1}. We call the mirror A at $\mathbf{X}_1$ the recycling mirror, the mirror B at $\mathbf{X}_1$ the steering mirror and the mirrors at $\mathbf{X}_2$ and $\mathbf{X}_3$ the end mirrors. The amplitude reflectivities and transmissivities of the steering mirror and the end mirrors at $\mathbf{X}_2$ and $\mathbf{X}_3$ are $(R_B,T_B)$, $(R_2,T_2)$, $(R_3,T_3)$, respectively. Those of the recycling mirror are $(+R_A, +T_A)$ for the light incident from inside the cavity and $(-R_A, +T_A)$ for the light incident from outside the cavity. The angular frequency of light is $\omega$ and the arm length is $L$. We define $\tau \equiv L/c$. Electric fields at the recycling mirror are defined in Fig.\ref{fig11} and are related by the following equations,
\begin{eqnarray}
C_{\ell}&=&R_A D_{\ell}+T_A A_{\ell} \;, \\
\label{eq51}
C_{r}&=&R_A D_{r}+T_A A_{r} \;, \\
B_{\ell}&=&T_A D_{\ell}-R_A A_{\ell} \;, \\
B_{r}&=&T_A D_{r}-R_A A_{r} \;. 
\end{eqnarray}
We assume that there is no displacement noise, for example, thermal noise, seismic noise, radiation pressure noise etc., at $100 \, \rm{MHz}$. Then, D field after circulating the cavity experiences the phase shift $4\omega \tau$ and the GW signal $\delta \phi (t)$,
\begin{eqnarray}
D_{\ell}(t)&=&R_c C_{\ell}(t-4\tau)\;e^{i [ 4\omega \tau +\delta \phi _{\ell}(t)]} \;, \\
D_{r}(t)&=&R_c C_{r}(t-4\tau)\;e^{i[4\omega \tau +\delta \phi _{r}(t)]}\;, 
\label{eq52}
\end{eqnarray}
where we defined the composite reflectivity of mirrors $R_c \equiv R_B R_2 R_3$ and 
\begin{eqnarray}
\delta \phi _{\ell}(t)&\equiv &\delta \phi _{21}(t)+\delta \phi _{12}(t-\tau)+\delta \phi _{31}(t-2\tau)+\delta \phi _{13}(t-3\tau) \;,  \label{eq54} \\
\delta \phi _{r}(t)&\equiv &\delta \phi _{31}(t)+\delta \phi _{13}(t-\tau)+\delta \phi _{21}(t-2\tau)+\delta \phi _{12}(t-3\tau)\;.
\end{eqnarray}
For example, $\tilde{\delta \phi}_{21}$ denotes the phase shift due to GW when light travels from $\mathbf{X}_2$ to $\mathbf{X}_1$ in Fig.\ref{fig1}. This can be calculated using Eq. (\ref{eq1}). 

Equations (\ref{eq51})-(\ref{eq52}) can be solved separately for right-handed and left-handed fields. The input-output relation of the cavity for the left-handed fields becomes
\begin{equation}
B_{\ell}(t)=-R_A A_{\ell}(t)+\sum_{k=1}^{\infty} T_A^2 R_c^k R_A^{k-1} A_{\ell}(t-4k\tau )e^{4i\omega \tau}\; \exp \left[ i\sum_{k^{\prime}=1}^{k} \delta \phi _{\ell} (t-4(k^{\prime}-1)\tau) \right]\;. \nonumber
\end{equation} 
We assume that the cavity is in resonance in the absence of GW, that is $A_{\ell}(t)=A_{\ell}(t-4\tau)$. Then,
\begin{equation}
B_{\ell}(t)=-R_A A_{\ell}(t) \left\{  1-\sum_{k=1}^{\infty} T_A^2 R_c^k R_A^{k-2}\; \exp\left[ i \sum_{k^{\prime}=1}^{k} \delta \phi _{\ell} (t-4(k^{\prime}-1)\tau ) \right]  \right\} \;. \nonumber 
\end{equation}
Using the approximation that the GW signal is small ($\delta \phi (t) \ll 1$), we obtain
\begin{equation}
B_{\ell}(t) \approx -\frac{ R_A- R_c}{1-R_A R_c}A_{\ell}(t)\; \exp \left[ -i \;\frac{1-R_A R_c}{R_A - R_c}\sum_{k=1}^{\infty} T_A^2 R_c^k R_A^{k-1}\;  \sum_{k^{\prime}=1}^{k} \delta \phi _{\ell} (t-4(k^{\prime}-1)\tau ) \right] \;. 
\label{eq69}
\end{equation}
Therefore, the phase shift $\delta \Phi_{\ell}$ of left-handed light caused by GW is
\begin{equation}
\delta \Phi _{\ell}(t) =-\frac{1-R_A R_c}{R_A -  R_c}\sum_{k=1}^{\infty} T_A^2 R_c^k R_A^{k-1}\;  \sum_{k^{\prime}=1}^{k} \delta \phi _{\ell} (t-4(k^{\prime}-1)\tau ) \;. 
\label{eq53}
\end{equation} 
Fourier transforming Eq. (\ref{eq53}) and using Eqs. (\ref{eq54}) and (\ref{eq2}) give
\begin{eqnarray}
\tilde{\delta \Phi }_{\ell}&=&\alpha (\Omega, R_c,R_{A}) \tilde{\delta \phi}_{\ell} \;, \\
\tilde{\delta \phi}_{\ell} &=& \frac{\omega}{i\Omega}\;e^{-i\Omega (\tau+\mathbf{e}_z \cdot \mathbf{X}_1/c)} \sum_p \mathbf{e}^p \tilde{h}_p \nonumber \\
&&: \left[ \right. \frac{\mathbf{n}_{12} \otimes \mathbf{n}_{12}}{1-(\mathbf{e}_z \cdot \mathbf{n}_{12})^2} \{ i \;\sin\Omega \tau +(\mathbf{e}_z \cdot \mathbf{n}_{12})(e^{-i\Omega \tau \mathbf{e}_z \cdot \mathbf{n}_{12}}-\cos\Omega \tau) \} \nonumber \\
&&\;+\frac{\mathbf{n}_{13} \otimes \mathbf{n}_{13}}{1-(\mathbf{e}_z \cdot \mathbf{n}_{13})^2}\;e^{-2i\Omega \tau} \{ i \;\sin\Omega \tau +(\mathbf{e}_z \cdot \mathbf{n}_{13})(e^{-i\Omega \tau \mathbf{e}_z \cdot \mathbf{n}_{13}}-\cos\Omega \tau) \} \left. \right] \;, \nonumber \\
&&  \\ 
\alpha (\Omega, R_A, R_c) &\equiv &-\frac{T_A^2 R_c}{(R_{A}-R_c)(1-R_{A}R_c\;e^{-4i\Omega \tau})}\;. \\
&& \nonumber 
\end{eqnarray}

The GW signal for the right-handed light can be obtained by simply changing the subscripts $2\leftrightarrow 3$ because of the symmetry of the system. Therefore, the output of the detector is 
\begin{eqnarray}
\tilde{\delta \Phi}_{all} &\equiv &\tilde{\delta \Phi }_{r}-\tilde{\delta \Phi }_{\ell} \nonumber \\
&=& \alpha (\Omega, R_A, R_c) \tilde{\delta \phi}_{all} \\
\tilde{\delta \phi}_{all} &=& (1-e^{-2i\Omega \tau}) \frac{\omega}{i\Omega}\;e^{-i\Omega (\tau+\mathbf{e}_z \cdot \mathbf{X}_1/c)} \sum_p \mathbf{e}^p \tilde{h}_p \nonumber \\
&&: \left[ \right. \frac{\mathbf{n}_{13} \otimes \mathbf{n}_{13}}{1-(\mathbf{e}_z \cdot \mathbf{n}_{13})^2} \{ i \;\sin\Omega \tau +(\mathbf{e}_z \cdot \mathbf{n}_{13})(e^{-i\Omega \tau \mathbf{e}_z \cdot \mathbf{n}_{13}}-\cos\Omega \tau) \} \nonumber \\
&&\;-\frac{\mathbf{n}_{12} \otimes \mathbf{n}_{12}}{1-(\mathbf{e}_z \cdot \mathbf{n}_{12})^2} \{ i \;\sin\Omega \tau +(\mathbf{e}_z \cdot \mathbf{n}_{12})(e^{-i\Omega \tau \mathbf{e}_z \cdot \mathbf{n}_{12}}-\cos\Omega \tau) \} \left. \right] \;.
\label{eq63}
\end{eqnarray}

\section{Response function of FPMI}
\begin{figure}[h]
\begin{center}
\includegraphics[width=8cm]{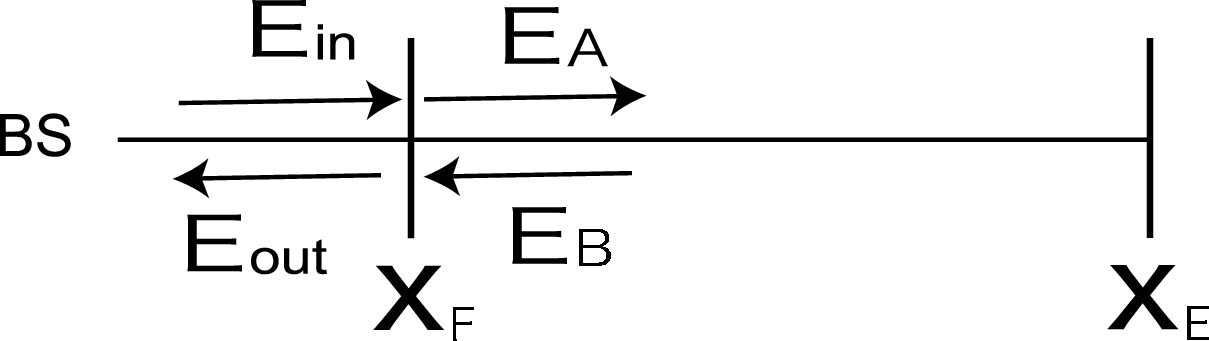}
\caption{Electric fields of FPMI.}
\label{fig12}
\end{center}
\end{figure}

The configuration of FPMI is shown in Fig.\ref{fig2}. The amplitude reflectivities and transmissivities of end mirrors at $\mathbf{X}_{E1}$ and $\mathbf{X}_{E2}$ are 
$(R_E,T_E)$, and of front mirrors at $\mathbf{X}_{F1}$ and $\mathbf{X}_{F2}$ are $(+R_F, +T_F)$ for the light incident from inside the cavities and $(-R_F, +T_F)$ for the light incident from outside the cavities. The arm length is $L$. Electric fields at the front mirror are defined in Fig.\ref{fig12}. First, we will consider only one FP cavity and calculate the input-output relation. At the end of our calculation, we will derive the full output of FPMI. 

The fields are related by the following equations,
\begin{eqnarray}
E_{out}&=&-R_F E_{in} +T_F E_B \label{eq55} \;, \\
E_A&=&R_F E_B +T_F E_{in} \;.\label{eq59}
\end{eqnarray}
$E_A$ is reflected at the end mirror and returns to the front mirror after experiencing the phase shift $2\tau$ and the modulation due to GW. The relation between $E_A$ and $E_B$ is
\begin{eqnarray}
E_B(t)&=&R_E E_A(t-2\tau)\;e^{i[2\omega \tau+\delta \phi _{cav}(t) ]} \;,
\label{eq56} \\
\delta \phi _{cav}(t)&\equiv &\delta \phi_{EF}(t)+\delta \phi_{FE}(t-\tau)\;,
\label{eq58}
\end{eqnarray}
where $\delta \phi_{EF}$ denotes the phase shift due to GW when light propagates from the end mirror to the front mirror and the specific form is given by Eq. (\ref{eq1}). Similarly, $\delta \phi_{FE}$ is the one from the front mirror to the end mirror. From Eqs.(\ref{eq55})-(\ref{eq56}), the input-output relation for the FP cavity becomes
\begin{eqnarray}
E_{out}(t)&=& -R_F E_{in}(t)+R_E T_F^2 \sum_{k=1}^{\infty} (R_F R_E)^{k-1} E_{in}(t) \;\exp[i\sum_{k^{\prime}=1}^{k}\delta \phi_{cav} (t-2(k^{\prime}-1)\tau )] \nonumber \\
&\approx & - \frac{R_F-R_E}{1-R_F R_E}E_{in}(t) \; \exp \left[ -i\frac{R_E T_F^2(1-R_F R_E)}{R_F-R_E} \sum_{k=1}^{\infty} (R_F R_E)^{k-1} \sum_{k^{\prime}=1}^{k}\delta \phi_{cav} (t-2(k^{\prime}-1)\tau ) \right] \;, \nonumber 
\end{eqnarray}
where we used the approximation $\delta \phi _{cav}(t) \ll 1$. Therefore, the phase shift $\delta \Phi$ caused by GW is
\begin{equation}
\delta \Phi (t)=-\frac{R_E T_F^2(1-R_F R_E)}{R_F-R_E} \sum_{k=1}^{\infty} (R_F R_E)^{k-1} \sum_{k^{\prime}=1}^{k}\delta \phi_{cav} (t-2(k^{\prime}-1)\tau ) \;. 
\label{eq57}
\end{equation}
Fourier transforming Eq. (\ref{eq57}) and using Eqs. (\ref{eq58}) and (\ref{eq2}) gives 
\begin{eqnarray}
\tilde{\Phi}(\Omega) &=& \alpha (\Omega, R_F,R_E) \tilde{\delta \phi_{cav}}(\Omega ) \;, \\
\tilde{\delta \phi}_{cav} &=& \frac{\omega}{i\Omega} \sum_{p} \mathbf{e}^p \tilde{h}_p \;e^{-i\Omega (\tau+\mathbf{e}_z \cdot \mathbf{X}_F/c)} \frac{\mathbf{n} \otimes \mathbf{n}}{ 1-(\mathbf{e}_z \cdot \mathbf{n})^2 }  \nonumber \\
&\times &\left[ \right.  i\;\sin\Omega \tau +(\mathbf{e}_z \cdot \mathbf{n})(e^{-i\Omega \tau \mathbf{e}_z \cdot \mathbf{n}}-\cos\Omega \tau) \left. \right]\;, \\
\alpha (\Omega, R_F,R_E) &\equiv &-\frac{R_E T_F^2}{(R_F-R_E)(1-R_F R_E \;e^{-2i\Omega \tau})} \;,
\end{eqnarray}
where $\mathbf{n} \equiv (\mathbf{X}_E-\mathbf{X}_F)/L$. This formula is consistent with the previous result, Eq.(6) in \cite{bib28}, except for an overall constant factor.

Using the result obtained above, we can easily obtain the full-output signal for the FPMI. We fix the subscripts $1$ and $2$ to distinguish north and east arms in Fig.\ref{fig2}, and define $\mathbf{n}_i \equiv (\mathbf{X}_{Ei}-\mathbf{X}_{Fi})/L,\;i=1,2$. For simplicity, we assume that the two front mirrors on both arms are located at the same place, that is, $\mathbf{X}_F=\mathbf{X}_{F1}=\mathbf{X}_{F2}$. This assumption is valid because it hardly affects the GW signal. Then, total output of FPMI is
\begin{eqnarray}
\tilde{\delta \Phi}_{all} &\equiv &  \tilde{\delta \Phi}_1-\tilde{\delta \Phi}_2 \nonumber \\
&=& \alpha (\Omega, R_F,R_E) \tilde{\delta \phi}_{all} \\
\tilde{\delta \phi}_{all}&=& \frac{\omega}{i\Omega} \sum_{p} \mathbf{e}^p \tilde{h}_p \;e^{-i\Omega (\tau+\mathbf{e}_z \cdot \mathbf{X}_F/c)} \nonumber \\
&&: \left[ \right. \frac{\mathbf{n}_1 \otimes \mathbf{n}_1}{ 1-(\mathbf{e}_z \cdot \mathbf{n}_1)^2 } 
\left\{ \right. i\;\sin\Omega \tau +(\mathbf{e}_z \cdot \mathbf{n}_1)(e^{-i\Omega \tau \mathbf{e}_z \cdot \mathbf{n}_1}-\cos\Omega \tau) \left. \right\} \nonumber \\
&&-\frac{\mathbf{n}_2 \otimes \mathbf{n}_2}{ 1-(\mathbf{e}_z \cdot \mathbf{n}_2)^2 } 
\left\{ \right.  i\;\sin\Omega \tau +(\mathbf{e}_z \cdot \mathbf{n}_2)(e^{-i\Omega \tau \mathbf{e}_z \cdot \mathbf{n}_2}-\cos\Omega \tau) \left. \right\} \left. \right] \;.
\label{eq64}
\end{eqnarray}

\section{Response function of LMI}
\begin{figure}[h]
\begin{center}
\includegraphics[width=5cm]{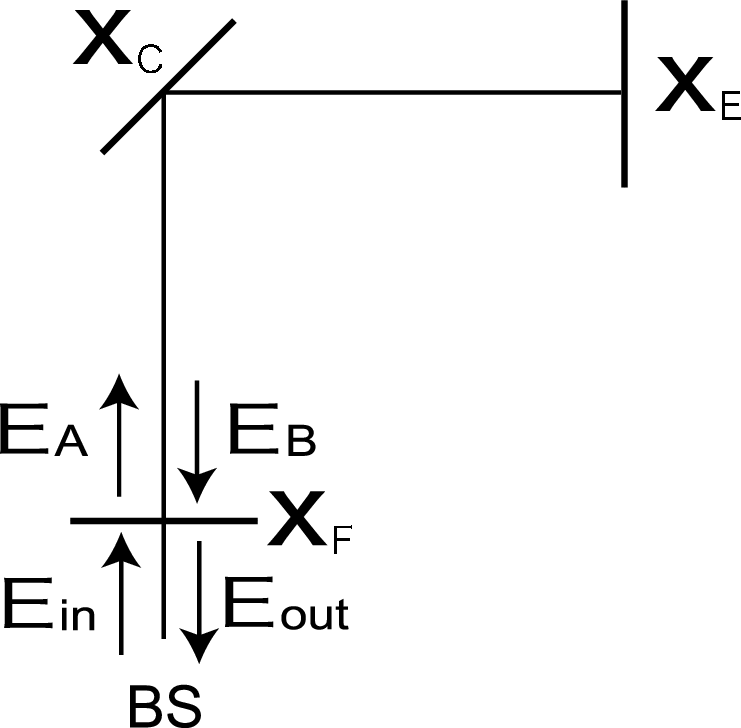}
\caption{Electric fields of LMI.}
\label{fig13}
\end{center}
\end{figure}

The configuration of LMI is shown in Fig.\ref{fig3}. The amplitude reflectivities and transmissivities of end mirrors at $\mathbf{X}_{E1}$ and $\mathbf{X}_{E2}$ are 
$(R_E,T_E)$, and of front mirrors at $\mathbf{X}_{F1}$ and $\mathbf{X}_{F2}$ are $(+R_F, +T_F)$ for the light incident from inside the cavities and $(-R_F, +T_F)$ for the light incident from outside the cavities. The two mirrors at $\mathbf{X}_{C1}$ and $\mathbf{X}_{C2}$ are completely reflective. The arm length is $L$. Electric fields at the front mirror are defined in Fig.\ref{fig13}. First, we will consider only one L-shaped cavity and calculate the input-output relation. At the end of our calculation, we will derive the full output of LMI. 

The relation between the fields is the same as Eqs. (\ref{eq55}) and (\ref{eq59}) for FPMI. The relation between $E_B$ and $E_A$ is almost the same as FPMI. However, only differences are that the round-trip time of LMI in the cavity is $4\tau$ and that light is reflected by the mirror at $\mathbf{X}_c$ during the trip. Thus, it is given by
\begin{eqnarray}
E_B(t)&=&R_E E_A(t-4\tau)\;e^{i[4\omega \tau+\delta \phi _{cav}(t) ]} \;, \label{eq60} \\
\delta \phi _{cav} (t) &\equiv &\delta \phi_{CF}(t)+\delta \phi_{EC}(t-\tau)+\delta \phi_{CE}(t-2\tau)+\delta \phi_{FC}(t-3\tau)\;,
\label{eq62}
\end{eqnarray}
where $\delta \phi_{IJ},\;I,J=F,E,C$ denotes the phase shift due to GW when the light propagates from the I-th mirror to the J-th mirror. From Eqs. (\ref{eq55}), (\ref{eq59}) and (\ref{eq60}), the input-output relation for the L-shaped cavity becomes
\begin{eqnarray}
E_{out}(t)&=& -R_F E_{in}(t)+R_E T_F^2 \sum_{k=1}^{\infty} (R_F R_E)^{k-1} E_{in}(t) \; \exp[i\sum_{k^{\prime}=1}^{k} \delta \phi_{cav} (t-4(k^{\prime}-1)\tau )] \nonumber \\
&\approx & - \frac{R_F-R_E}{1-R_F R_E}E_{in}(t) \; \exp \left[ -i\frac{R_E T_F^2(1-R_F R_E)}{R_F-R_E} \sum_{k=1}^{\infty} (R_F R_E)^{k-1} \sum_{k^{\prime}=1}^{k} \delta \phi_{cav} (t-4(k^{\prime}-1)\tau ) \right] \;, \nonumber 
\end{eqnarray}
where we used the approximation $\delta \phi _{cav}(t) \ll 1$. Therefore, the phase shift $\delta \Phi$ caused by GW is
\begin{equation}
\delta \Phi (t)=-\frac{R_E T_F^2(1-R_F R_E)}{R_F-R_E} \sum_{k=1}^{\infty} (R_F R_E)^{k-1} \sum_{k^{\prime}=1}^{k}\delta \phi_{cav} (t-4(k^{\prime}-1)\tau ) \;,
\label{eq61}
\end{equation}
Fourier transforming Eq. (\ref{eq61}) and using Eqs. (\ref{eq62}) and (\ref{eq2}) gives 
\begin{eqnarray}
\tilde{\delta \Phi}(\Omega) &=& \alpha (\Omega, R_F,R_E) \tilde{\delta \phi_{cav}}(\Omega ) \;,\\
\tilde{\delta \phi}_{cav} &=& \frac{\omega}{i\Omega} \sum_{p} \mathbf{e}^p \tilde{h}_p \;e^{-i\Omega (2\tau+\mathbf{e}_z \cdot \mathbf{X}_C/c)} \nonumber \\
&\times &\left[ \right. \frac{\mathbf{n}_{CE}\otimes \mathbf{n}_{CE}}{ 1-(\mathbf{e}_z \cdot \mathbf{n}_{CE})^2 } \left\{ i\;\sin\Omega \tau +(\mathbf{e}_z \cdot \mathbf{n}_{CE})(e^{-i\Omega \tau \mathbf{e}_z \cdot \mathbf{n}_{CE} }-\cos\Omega \tau) \right\} \nonumber \\
&&-\frac{\mathbf{n}_{CF}\otimes \mathbf{n}_{CF}}{ 1-(\mathbf{e}_z \cdot \mathbf{n}_{CF})^2 } \left\{ \right. i\;\sin\Omega \tau(1-2e^{-i\Omega \tau \mathbf{e}_z \cdot \mathbf{n}_{CF} }\cos\Omega \tau) \nonumber \\
&&\;\;\;\;\;\;\;\;\;\;\;\;\;\;\;\;\;\;\;\;\;\;\;\;\;\;\;\;\;\;-(\mathbf{e}_z \cdot \mathbf{n}_{CF})(e^{-i\Omega \tau \mathbf{e}_z \cdot \mathbf{n}_{CF} }\cos2\Omega \tau-\cos\Omega \tau) \left. \right\} \left. \right] \;, \nonumber \\
&& \\
\alpha (\Omega, R_F,R_E) &\equiv &-\frac{R_E T_F^2}{(R_F-R_E)(1-R_F R_E \;e^{-4i\Omega \tau})}\;,
\end{eqnarray}
where $\mathbf{n}_{CE} \equiv (\mathbf{X}_E-\mathbf{X}_C)/L$A$\mathbf{n}_{CF} \equiv (\mathbf{X}_F-\mathbf{X}_C)/L$.

Using the result obtained above, we can easily obtain the full-output signal for the LMI. We fix the subscripts $1$ and $2$ in order to distinguish the two arms in Fig.\ref{fig3}, and define $\mathbf{n}_{1} \equiv \mathbf{n}_{C2E}=-\mathbf{n}_{C1F}$ and $\mathbf{n}_{2} \equiv -\mathbf{n}_{C2F}=\mathbf{n}_{C1E}$. For simplicity, we assume that the two front mirrors and the two end mirrors on both arms are located at the same place, that is, $\mathbf{X}_F=\mathbf{X}_{F1}=\mathbf{X}_{F2}$ and $\mathbf{X}_E=\mathbf{X}_{E1}=\mathbf{X}_{E2}$, respectively. This assumption is valid because it hardly affects the GW signal. Then, total output of LMI is
\begin{eqnarray}
\tilde{\delta \Phi}_{all} &\equiv & \tilde{\delta \Phi}_1-\tilde{\delta \Phi}_2 \nonumber \\
&=& \alpha (\Omega, R_F,R_E) \tilde{\delta \phi}_{all} \\
\tilde{\delta \phi}_{all} &=& \frac{\omega}{i\Omega} \sum_{p} \mathbf{e}^p \tilde{h}_p \;e^{-i\Omega (2\tau+\mathbf{e}_z \cdot \mathbf{X}_{F}/c)} \nonumber \\
&: &\left[ \right. \frac{\mathbf{n}_{2}\otimes \mathbf{n}_{2}}{ 1-(\mathbf{e}_z \cdot \mathbf{n}_{2})^2 } \left\{ \right. i\;\sin\Omega \tau\; (e^{-ip_1}+e^{-ip_2}-2\cos\Omega \tau) \nonumber \\
&&\;\;\;\;\;\;\;\;\;\;\;\;\;\;\;\;\;\;\;\;\;\;\;\;\;\;\;+(\mathbf{e}_z \cdot \mathbf{n}_{2})(e^{-i(p_1+p_2)}+\cos2\Omega L -(e^{-ip_1}+e^{-ip_2})\cos\Omega \tau) \left. \right\} \nonumber \\
&&-\frac{\mathbf{n}_{1}\otimes \mathbf{n}_{1}}{ 1-(\mathbf{e}_z \cdot \mathbf{n}_{1})^2 } \left\{ \right. i\;\sin\Omega \tau\; (e^{-ip_1}+e^{-ip_2}-2\cos\Omega \tau) \nonumber \\
&&\;\;\;\;\;\;\;\;\;\;\;\;\;\;\;\;\;\;\;\;\;\;\;\;\;\;\;+(\mathbf{e}_z \cdot \mathbf{n}_{1})(e^{-i(p_1+p_2)}+\cos2\Omega \tau -(e^{-ip_1}+e^{-ip_2})\cos\Omega \tau) \left. \right\} \left. \right] \;,\nonumber \\
&& \label{eq65} 
\end{eqnarray}
where we defined $p_1 \equiv \Omega \tau (\mathbf{e}_z \cdot \mathbf{n}_1)$ and $p_2 \equiv \Omega \tau (\mathbf{e}_z \cdot \mathbf{n}_2)$.

\bibliography{UHFGW1}

\begin{thebibliography}{32}
\expandafter\ifx\csname natexlab\endcsname\relax\def\natexlab#1{#1}\fi
\expandafter\ifx\csname bibnamefont\endcsname\relax
  \def\bibnamefont#1{#1}\fi
\expandafter\ifx\csname bibfnamefont\endcsname\relax
  \def\bibfnamefont#1{#1}\fi
\expandafter\ifx\csname citenamefont\endcsname\relax
  \def\citenamefont#1{#1}\fi
\expandafter\ifx\csname url\endcsname\relax
  \def\url#1{\texttt{#1}}\fi
\expandafter\ifx\csname urlprefix\endcsname\relax\def\urlprefix{URL }\fi
\providecommand{\bibinfo}[2]{#2}
\providecommand{\eprint}[2][]{\url{#2}}

\bibitem[{\citenamefont{J.~E.~Peebles and Vilenkin}(1999)}]{bib10}
\bibinfo{author}{\bibfnamefont{P.}~\bibnamefont{J.~E.~Peebles}}
  \bibnamefont{and} \bibinfo{author}{\bibfnamefont{A.}~\bibnamefont{Vilenkin}},
  \bibinfo{journal}{Phys.\ Rev.} \textbf{\bibinfo{volume}{D 59}},
  \bibinfo{pages}{063505} (\bibinfo{year}{1999}).

\bibitem[{\citenamefont{Giovannini}(1999)}]{bib11}
\bibinfo{author}{\bibfnamefont{M.}~\bibnamefont{Giovannini}},
  \bibinfo{journal}{Phys.\ Rev.} \textbf{\bibinfo{volume}{D 60}},
  \bibinfo{pages}{123511} (\bibinfo{year}{1999}).

\bibitem[{\citenamefont{Giovannini}(1998)}]{bib12}
\bibinfo{author}{\bibfnamefont{M.}~\bibnamefont{Giovannini}},
  \bibinfo{journal}{Phys.\ Rev.} \textbf{\bibinfo{volume}{D 58}},
  \bibinfo{pages}{083504} (\bibinfo{year}{1998}).

\bibitem[{\citenamefont{Riazuelo and Uzan}(2000)}]{bib13}
\bibinfo{author}{\bibfnamefont{A.}~\bibnamefont{Riazuelo}} \bibnamefont{and}
  \bibinfo{author}{\bibfnamefont{J.~P.} \bibnamefont{Uzan}},
  \bibinfo{journal}{Phys.\ Rev.} \textbf{\bibinfo{volume}{D 62}},
  \bibinfo{pages}{083506} (\bibinfo{year}{2000}).

\bibitem[{\citenamefont{Tashiro et~al.}(2004)\citenamefont{Tashiro, Chiba, and
  Sasaki}}]{bib14}
\bibinfo{author}{\bibfnamefont{H.}~\bibnamefont{Tashiro}},
  \bibinfo{author}{\bibfnamefont{T.}~\bibnamefont{Chiba}}, \bibnamefont{and}
  \bibinfo{author}{\bibfnamefont{M.}~\bibnamefont{Sasaki}},
  \bibinfo{journal}{Class.\ Quantum.\ Grav.} \textbf{\bibinfo{volume}{21}},
  \bibinfo{pages}{1761} (\bibinfo{year}{2004}).

\bibitem[{\citenamefont{Easther and Lim}()}]{bib29}
\bibinfo{author}{\bibfnamefont{R.}~\bibnamefont{Easther}} \bibnamefont{and}
  \bibinfo{author}{\bibfnamefont{E.~A.} \bibnamefont{Lim}},
  \bibinfo{note}{astro-ph/0601617}.

\bibitem[{\citenamefont{Garcia-Bellido and Figueroa}(2007)}]{bib30}
\bibinfo{author}{\bibfnamefont{J.}~\bibnamefont{Garcia-Bellido}}
  \bibnamefont{and} \bibinfo{author}{\bibfnamefont{D.~G.}
  \bibnamefont{Figueroa}}, \bibinfo{journal}{Phys.\ Rev.\ Lett.}
  \textbf{\bibinfo{volume}{98}}, \bibinfo{pages}{061302}
  (\bibinfo{year}{2007}).

\bibitem[{\citenamefont{Dufaux et~al.}()\citenamefont{Dufaux, Bergman, Felder,
  Kofman, and Uzan}}]{bib34}
\bibinfo{author}{\bibfnamefont{J.}~\bibnamefont{Dufaux}},
  \bibinfo{author}{\bibfnamefont{A.}~\bibnamefont{Bergman}},
  \bibinfo{author}{\bibfnamefont{G.}~\bibnamefont{Felder}},
  \bibinfo{author}{\bibfnamefont{L.}~\bibnamefont{Kofman}}, \bibnamefont{and}
  \bibinfo{author}{\bibfnamefont{J.}~\bibnamefont{Uzan}},
  \bibinfo{note}{arXiv:0707.0875}.

\bibitem[{\citenamefont{Gasperini and Giovannini}(1993)}]{bib15}
\bibinfo{author}{\bibfnamefont{M.}~\bibnamefont{Gasperini}} \bibnamefont{and}
  \bibinfo{author}{\bibfnamefont{M.}~\bibnamefont{Giovannini}},
  \bibinfo{journal}{Phys.\ Rev.} \textbf{\bibinfo{volume}{D 47}},
  \bibinfo{pages}{1519} (\bibinfo{year}{1993}).

\bibitem[{\citenamefont{Brustein et~al.}(1995)\citenamefont{Brustein,
  Gasperini, Giovannini, and Veneziano}}]{bib16}
\bibinfo{author}{\bibfnamefont{R.}~\bibnamefont{Brustein}},
  \bibinfo{author}{\bibfnamefont{M.}~\bibnamefont{Gasperini}},
  \bibinfo{author}{\bibfnamefont{M.}~\bibnamefont{Giovannini}},
  \bibnamefont{and}
  \bibinfo{author}{\bibfnamefont{G.}~\bibnamefont{Veneziano}},
  \bibinfo{journal}{Phys.\ Lett.} \textbf{\bibinfo{volume}{B 361}},
  \bibinfo{pages}{45} (\bibinfo{year}{1995}).

\bibitem[{\citenamefont{Gasperini and Veneziano}(2003)}]{bib17}
\bibinfo{author}{\bibfnamefont{M.}~\bibnamefont{Gasperini}} \bibnamefont{and}
  \bibinfo{author}{\bibfnamefont{G.}~\bibnamefont{Veneziano}},
  \bibinfo{journal}{Phys.\ Rep.} \textbf{\bibinfo{volume}{373}},
  \bibinfo{pages}{1} (\bibinfo{year}{2003}).

\bibitem[{\citenamefont{Nakamura et~al.}(1997)\citenamefont{Nakamura, Sasaki,
  Tanaka, and Thorne}}]{bib18}
\bibinfo{author}{\bibfnamefont{T.}~\bibnamefont{Nakamura}},
  \bibinfo{author}{\bibfnamefont{M.}~\bibnamefont{Sasaki}},
  \bibinfo{author}{\bibfnamefont{T.}~\bibnamefont{Tanaka}}, \bibnamefont{and}
  \bibinfo{author}{\bibfnamefont{K.~S.} \bibnamefont{Thorne}},
  \bibinfo{journal}{Astrophys.\ J.} \textbf{\bibinfo{volume}{487}},
  \bibinfo{pages}{L139} (\bibinfo{year}{1997}).

\bibitem[{\citenamefont{Ioka et~al.}(1998)\citenamefont{Ioka, Chiba, Tanaka,
  and Nakamura}}]{bib19}
\bibinfo{author}{\bibfnamefont{K.}~\bibnamefont{Ioka}},
  \bibinfo{author}{\bibfnamefont{T.}~\bibnamefont{Chiba}},
  \bibinfo{author}{\bibfnamefont{T.}~\bibnamefont{Tanaka}}, \bibnamefont{and}
  \bibinfo{author}{\bibfnamefont{T.}~\bibnamefont{Nakamura}},
  \bibinfo{journal}{Phys.\ Rev.} \textbf{\bibinfo{volume}{D 58}},
  \bibinfo{pages}{063003} (\bibinfo{year}{1998}).

\bibitem[{\citenamefont{S.~Bisnovatyi-Kogan and Rudenko}(2004)}]{bib20}
\bibinfo{author}{\bibfnamefont{G.}~\bibnamefont{S.~Bisnovatyi-Kogan}}
  \bibnamefont{and} \bibinfo{author}{\bibfnamefont{V.~N.}
  \bibnamefont{Rudenko}}, \bibinfo{journal}{Class.\ Quantum.\ Grav.}
  \textbf{\bibinfo{volume}{21}}, \bibinfo{pages}{3347} (\bibinfo{year}{2004}).

\bibitem[{\citenamefont{S.~Seahra et~al.}(2005)\citenamefont{S.~Seahra,
  Clarkson, and Maartens}}]{bib21}
\bibinfo{author}{\bibfnamefont{S.}~\bibnamefont{S.~Seahra}},
  \bibinfo{author}{\bibfnamefont{C.}~\bibnamefont{Clarkson}}, \bibnamefont{and}
  \bibinfo{author}{\bibfnamefont{R.}~\bibnamefont{Maartens}},
  \bibinfo{journal}{Phys.\ Rev.\ Lett.} \textbf{\bibinfo{volume}{94}},
  \bibinfo{pages}{121302} (\bibinfo{year}{2005}).

\bibitem[{\citenamefont{Clarkson and Seahra}(2007)}]{bib22}
\bibinfo{author}{\bibfnamefont{C.}~\bibnamefont{Clarkson}} \bibnamefont{and}
  \bibinfo{author}{\bibfnamefont{S.~S.} \bibnamefont{Seahra}},
  \bibinfo{journal}{Class.\ Quantum.\ Grav.} \textbf{\bibinfo{volume}{24}},
  \bibinfo{pages}{F33} (\bibinfo{year}{2007}).

\bibitem[{\citenamefont{L.~Smith et~al.}(2006)\citenamefont{L.~Smith,
  Kamionkowski, and Cooray}}]{bib23}
\bibinfo{author}{\bibfnamefont{T.}~\bibnamefont{L.~Smith}},
  \bibinfo{author}{\bibfnamefont{M.}~\bibnamefont{Kamionkowski}},
  \bibnamefont{and} \bibinfo{author}{\bibfnamefont{A.}~\bibnamefont{Cooray}},
  \bibinfo{journal}{Phys.\ Rev.} \textbf{\bibinfo{volume}{D 73}},
  \bibinfo{pages}{023504} (\bibinfo{year}{2006}).

\bibitem[{\citenamefont{Jenet et~al.}(2006)}]{bib24}
\bibinfo{author}{\bibfnamefont{F.~A.} \bibnamefont{Jenet}}
  \bibnamefont{et~al.}, \bibinfo{journal}{Astrophys.\ J.}
  \textbf{\bibinfo{volume}{653}}, \bibinfo{pages}{1571} (\bibinfo{year}{2006}).

\bibitem[{\citenamefont{Armstrong et~al.}(2003)\citenamefont{Armstrong, Iess,
  Tortora, and Bertotti}}]{bib33}
\bibinfo{author}{\bibfnamefont{J.~W.} \bibnamefont{Armstrong}},
  \bibinfo{author}{\bibfnamefont{L.}~\bibnamefont{Iess}},
  \bibinfo{author}{\bibfnamefont{P.}~\bibnamefont{Tortora}}, \bibnamefont{and}
  \bibinfo{author}{\bibfnamefont{B.}~\bibnamefont{Bertotti}},
  \bibinfo{journal}{Astrophys.\ J.} \textbf{\bibinfo{volume}{599}},
  \bibinfo{pages}{806} (\bibinfo{year}{2003}).

\bibitem[{\citenamefont{Abbot et~al.}(2007)}]{bib25}
\bibinfo{author}{\bibfnamefont{B.}~\bibnamefont{Abbot}} \bibnamefont{et~al.},
  \bibinfo{journal}{Astrophys.\ J.} \textbf{\bibinfo{volume}{659}},
  \bibinfo{pages}{918} (\bibinfo{year}{2007}).

\bibitem[{\citenamefont{Maggiore}(2000)}]{bib26}
\bibinfo{author}{\bibfnamefont{M.}~\bibnamefont{Maggiore}},
  \bibinfo{journal}{Phys.\ Rep.} \textbf{\bibinfo{volume}{331}},
  \bibinfo{pages}{283} (\bibinfo{year}{2000}).

\bibitem[{\citenamefont{M.~Cruise and Ingley}(2006)}]{bib27}
\bibinfo{author}{\bibfnamefont{A.}~\bibnamefont{M.~Cruise}} \bibnamefont{and}
  \bibinfo{author}{\bibfnamefont{R.~M.~J.} \bibnamefont{Ingley}},
  \bibinfo{journal}{Class.\ Quantum.\ Grav.} \textbf{\bibinfo{volume}{23}},
  \bibinfo{pages}{6185} (\bibinfo{year}{2006}).

\bibitem[{\citenamefont{W.~P.~Drever}()}]{bib1}
\bibinfo{author}{\bibfnamefont{R.}~\bibnamefont{W.~P.~Drever}},
  \emph{\bibinfo{title}{Gravitational radiation}}, \bibinfo{note}{edited by N.
  Deruelle and T. Piran (North-Holland, Amsterdam, 1983), pp.321-338}.

\bibitem[{\citenamefont{Vinet et~al.}(1988)\citenamefont{Vinet, Meers, Man, and
  Brillet}}]{bib31}
\bibinfo{author}{\bibfnamefont{J.~Y.} \bibnamefont{Vinet}},
  \bibinfo{author}{\bibfnamefont{B.~J.} \bibnamefont{Meers}},
  \bibinfo{author}{\bibfnamefont{C.~N.} \bibnamefont{Man}}, \bibnamefont{and}
  \bibinfo{author}{\bibfnamefont{A.}~\bibnamefont{Brillet}},
  \bibinfo{journal}{Phys.\ Rev.} \textbf{\bibinfo{volume}{D 38}},
  \bibinfo{pages}{433} (\bibinfo{year}{1988}).

\bibitem[{\citenamefont{Meers}(1988)}]{bib32}
\bibinfo{author}{\bibfnamefont{B.~J.} \bibnamefont{Meers}},
  \bibinfo{journal}{Phys.\ Rev.} \textbf{\bibinfo{volume}{D 38}},
  \bibinfo{pages}{2317} (\bibinfo{year}{1988}).

\bibitem[{\citenamefont{J.~Waldman et~al.}(2006)}]{bib2}
\bibinfo{author}{\bibfnamefont{S.}~\bibnamefont{J.~Waldman}}
  \bibnamefont{et~al.}, \bibinfo{journal}{Class.\ Quantum.\ Grav.}
  \textbf{\bibinfo{volume}{23}}, \bibinfo{pages}{S653} (\bibinfo{year}{2006}).

\bibitem[{\citenamefont{Acernese et~al.}(2006)}]{bib3}
\bibinfo{author}{\bibfnamefont{F.}~\bibnamefont{Acernese}}
  \bibnamefont{et~al.}, \bibinfo{journal}{Class.\ Quantum.\ Grav.}
  \textbf{\bibinfo{volume}{23}}, \bibinfo{pages}{S635} (\bibinfo{year}{2006}).

\bibitem[{\citenamefont{Hild et~al.}(2006)}]{bib4}
\bibinfo{author}{\bibfnamefont{S.}~\bibnamefont{Hild}} \bibnamefont{et~al.},
  \bibinfo{journal}{Class.\ Quantum.\ Grav.} \textbf{\bibinfo{volume}{23}},
  \bibinfo{pages}{S643} (\bibinfo{year}{2006}).

\bibitem[{\citenamefont{Ando et~al.}(2001)}]{bib5}
\bibinfo{author}{\bibfnamefont{M.}~\bibnamefont{Ando}} \bibnamefont{et~al.},
  \bibinfo{journal}{Phys.\ Rev.\ Lett.} \textbf{\bibinfo{volume}{86}},
  \bibinfo{pages}{3950} (\bibinfo{year}{2001}).

\bibitem[{\citenamefont{Chen and Kawamura}(2006)}]{bib6}
\bibinfo{author}{\bibfnamefont{Y.}~\bibnamefont{Chen}} \bibnamefont{and}
  \bibinfo{author}{\bibfnamefont{S.}~\bibnamefont{Kawamura}},
  \bibinfo{journal}{Phys.\ Rev.\ Lett.} \textbf{\bibinfo{volume}{96}},
  \bibinfo{pages}{231102} (\bibinfo{year}{2006}).

\bibitem[{\citenamefont{B.~Estabrook and Wahlquist}(1975)}]{bib7}
\bibinfo{author}{\bibfnamefont{F.}~\bibnamefont{B.~Estabrook}}
  \bibnamefont{and} \bibinfo{author}{\bibfnamefont{H.~D.}
  \bibnamefont{Wahlquist}}, \bibinfo{journal}{Gen.\ Relat.\ Gravit.}
  \textbf{\bibinfo{volume}{6}}, \bibinfo{pages}{439} (\bibinfo{year}{1975}).

\bibitem[{\citenamefont{Schilling}(1997)}]{bib28}
\bibinfo{author}{\bibfnamefont{R.}~\bibnamefont{Schilling}},
  \bibinfo{journal}{Class.\ Quantum.\ Grav.} \textbf{\bibinfo{volume}{14}},
  \bibinfo{pages}{1513} (\bibinfo{year}{1997}).

\end{thebibliography}

\end{document}